\documentclass[apj, numberedappendix]{openjournal}

\usepackage[dvipsnames]{xcolor}
\usepackage[breaklinks,colorlinks,linkcolor=BlueViolet,citecolor=BlueViolet,urlcolor=BlueViolet]{hyperref}
\usepackage{amsmath}
\usepackage[autostyle]{csquotes} 
\usepackage[T1]{fontenc} 
\usepackage{orcidlink} 
\usepackage{needspace} 

\makeatletter
\renewcommand{\frontmatter@title@above}{%
  \vspace*{-\headsep}\vspace*{\headheight}%
  \footnotesize
  {\footnotesize Submitted to The Open Journal of Astrophysics}\par 
  {\footnotesize\textsc{\@journalinfo}}\par 
  \vspace*{-\baselineskip}\vspace*{0.625in}%
}
\makeatother

\usepackage[english]{babel} 
\usepackage[autostyle]{csquotes}
\usepackage[T1]{fontenc} 

\newcommand{\rhigh}{R_{\text{high}}}
\newcommand{\rlow}{R_{\text{low}}}

\newcommand{\spin}{a_{\ast}}

\newcommand{\PA}{\mathrm{PA}}

\newcommand{\uv}{(u,v)}

\newcommand{\<}{\langle}
\renewcommand{\>}{\rangle}
\newcommand{\ipole}{{\tt ipole }}

\newcommand{\kharma}{{KHARMA }}
\newcommand{\patoka}{{\tt PATOKA }}

\shorttitle{Pitch Angles}
\shortauthors{Conroy et al.}

\begin{document}

\title{\vspace{-0.1cm}Spiral Pitch Angles in Black Hole Images and Movies\vspace{-1.5cm}}
	
\author{
	Nicholas S. Conroy\orcidlink{0000-0003-2886-2377},$^{1}$ 
	Michi Baub\"ock\orcidlink{0000-0002-5518-2812},$^{2,3}$ 
	Vedant Dhruv\orcidlink{0000-0001-6765-877X},$^{2,3,4}$
	Abhishek V. Joshi\orcidlink{0000-0002-2514-5965},$^{2,3}$\\
	Cora Prather\orcidlink{0000-0002-0393-7734},$^{5}$
	and Charles F. Gammie\orcidlink{0000-0001-7451-8935}$^{2,1,6,3}$
	\\
    {\scriptsize  \vspace{\baselineskip}
	$^{1}$Department of Astronomy, University of Illinois at Urbana-Champaign, 1002 West Green Street, Urbana, IL 61801, USA\\
	$^{2}$Department of Physics, University of Illinois at Urbana-Champaign, 1110 West Green Street, Urbana, IL 61801, USA\\ 
	$^{3}$Illinois Center for the Advanced Study of the Universe, University of Illinois at Urbana-Champaign,\\ 1110 West Green St., Urbana, IL 61801, USA\\
	$^{4}$Department of Physics and McDonnell Center for the Space Sciences, Washington University, St. Louis, MO 63130, USA\\
	$^{5}$Black Hole Initiative at Harvard University, 20 Garden Street, Cambridge, MA 02138, USA\\
	$^{6}$NCSA, University of Illinois at Urbana-Champaign, 1205 W. Clark St., Urbana, IL 61801, USA
    }
}
 
\begin{abstract}

The Event Horizon Telescope (EHT) has produced images of M87* and Sgr A*, and will soon produce time sequences of images, or movies. General relativistic magnetohydrodynamics (GRMHD) models predict EHT sources will exhibit brightness fluctuations along the ring. The fluctuations are commonly connected to trailing spiral features. Here, we introduce a method to measure the pitch angle $\psi$ of radially extended features by cross-correlating brightness fluctuations sampled on rings at two distinct impact parameters. Pitch angle can be reliably measured from synthetic movies with BHEX resolution, and can be constrained even at EHT resolution. Single images enable noisy constraints of $\psi$. Pitch angle depends most on the viewing inclination and black hole spin. For constant inclination, the spiral winding direction depends on the sign of spin, with spirals winding more tightly in the direction of frame dragging near the horizon. 
The pitch angle will complement other observables to reliably constrain both the source inclination and spin in future analysis by EHT, and by expansions to the EHT such as ngEHT and BHEX.

\end{abstract}

\maketitle

\section{Introduction}

The Event Horizon Telescope (EHT) has produced horizon-scale images of the black hole M87* at the center of M87 (\citealt{M87PaperI}) and the black hole Sagittarius A* (Sgr~A*) at the heart of our own galaxy (\citealt{SgrAPaperI}). These images were made by combining data from an array of radio telescopes using very long baseline interferometry (VLBI).  For M87*, the result is several snapshot images \citep{M87PaperI,M87PaperVII,M87_2018_paperI, M87_2021_paperI}. Each has observing duration comparable to the gravitational time $t_g\equiv G Mc^{-3} \sim 9 $ hrs, during which the source is expected to change relatively little.  For Sgr A*, the result is an approximate time-averaged image \citep{SgrAPaperI}, taken over many gravitational times $t_g \sim 20 \sec$. 

Ongoing observations will increase our sample size of horizon-scale images. Proposed expansions to EHT, such as the ngEHT \citep{Doeleman_2023_ngeht, Johnson_2023_ngeht, Ayzenberg_2025_ngEHT} and the Black Hole Explorer \citep[BHEX;][]{bhex_2024}, aim to capture horizon-scale images of new sources. 
Meanwhile, repeated observations, improved $\uv$ coverage, and improved imaging algorithms will enable movies (time sequences of images) for both M87* and Sgr A*. What information can be extracted from larger samples of images and movies, as they become feasible?

In a prior paper \citep{Conroy_2023}, we used state-of-the-art General Relativistic Magnetohydrodynamics (GRMHD) simulations to predict that dynamics in EHT movies will be dominated by apparent rotation. We measured this apparent rotation by correlating brightness temperature along the ring. The main result was that rotation was unexpectedly slow---slower than the Keplerian velocity of a feature just subject to gravitation, and slower than the fluid velocity (which can be arrested by pressure gradient and magnetic field forces). Rather than the fast motion of a hot spot, we predicted EHT movies will exhibit a pattern speed $\Omega_p$, driven by the slow azimuthal phase velocity of ingoing spiral waves. 

A central goal of the EHT is to measure black hole spin \citep{Midrange_science_goals}. We define the dimensionless spin $\spin \equiv Jc/GM^2$, for black hole angular momentum $J$, speed of light $c$, Newtonian gravitational constant $G$, and black hole mass $M$. Spin is one of the two parameters that describe an uncharged black hole. In an astronomical context, spin can influence the surrounding environment, powering relativistic jets \citep{Blandford_Znajek}, influencing the orbits of nearby stars \citep[e.g.][]{spin_stars1,spin_stars2,GRAVITY_2026_s301}, and influencing the behavior of the near-horizon accretion flow. Constraints on spin could offer insights into the black hole's formation, growth history, and host galaxy's merger history \citep[e.g.,][]{gammie_2004, Volonteri_2007_spin_host_morphology, Merloni_2008, bambi_2021, Reynolds_2021, Ricarte_2023, Ricarte_2025_spin_evolution}.
However, the pattern speed in movies depends most strongly on the viewing inclination $i$ relative to the accretion disk angular momentum, with only a weaker dependence on $\spin$. 

This motivated additional techniques to constrain spin. One promising technique uses polarization, as the linear polarization structure shows a clear spin dependence in models \citep[e.g.][]{Palumbo_2020, Chael_2023, Wong_2026, Chael_2026}. However, polarized observables are sensitive to additional uncertainties compared to intensity observables, including additional parameters (the magnetic field polarity), observational uncertainties in the compact flux, and intervening Faraday rotation. These uncertainties must be considered when performing polarization-based spin inference (R. Gunturu et al. in preparation). 

\begin{figure}[t]
    \centering
    \includegraphics[width=0.5\textwidth]{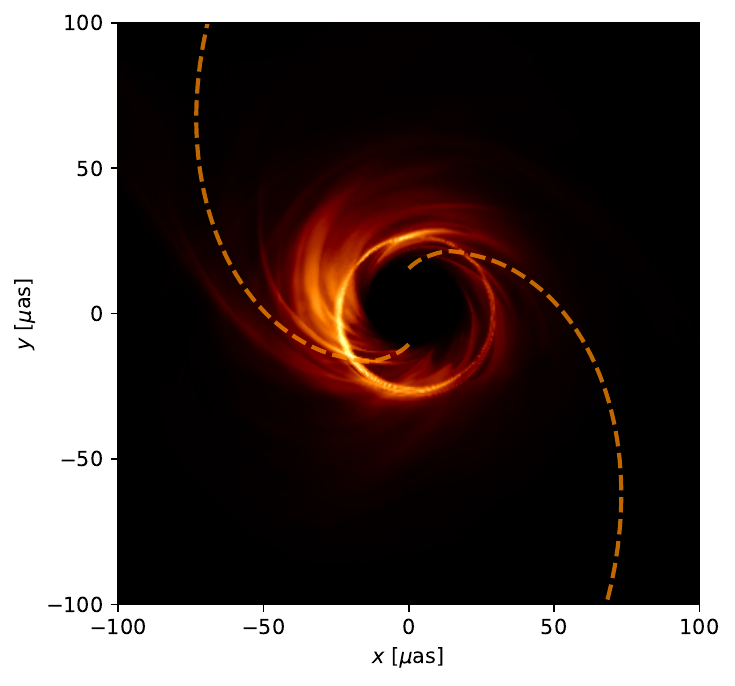}
    \caption{An example image from a fiducial Sgr A* GRMHD model (MAD, $\spin=0.5$, $i=30^\circ$, $\rhigh=160$). The two dashed orange curves depict the pitch angle $\psi$ measured in the video of this model at BHEX resolution.}
    \label{fig:frame}
\end{figure}

In intensity data, one promising technique uses the large-scale brightness asymmetry. The magnitude of the asymmetry typically increases with the magnitude of the spin, enabling statistically significant spin constraints \citep{Bernshteyn_2026}. However, these constraints are relatively insensitive to the sign of the spin (i.e. whether the black hole and its accretion disk are prograde or retrograde). One option to break this degeneracy combines the asymmetry magnitude and position angle (which depends on the spin vector position angle) with the pattern speed \citep{Conroy_2026_PA}. This may work well in M87*, where there exists a strong prior for inclination, but uncertainties in the inclination of Sgr A* and other, future sources may limit the precision of asymmetry-based spin inference, motivating complementary methods.

Here, we present a new technique that can constrain both the spin and inclination.
GRMHD models predict that the slow pattern speed is likely driven by ingoing spiral waves. It is natural to ask whether it is possible to measure the pitch angle of these spirals, along with other features, and whether the pitch angle depends on model parameters such as black hole spin.

The main contribution of this paper is to define a method for measuring the pitch angle $\psi$ in images and movies of EHT sources. We demonstrate that we can recover reasonable results even at the angular resolution of present-day EHT, and discuss how future ground-based expansions such as ngEHT can improve upon present constraints. We further analyze the dependence of $\psi$ on spin when measured with the high angular resolution of BHEX \citep{bhex_2024}.
An example frame from a fiducial GRMHD model, along with the measured pitch angle, is shown in Figure \ref{fig:frame}. Details on the GRMHD model library and methodology are found in the following sections. 
Our work builds on \cite{Ricarte_2022}, who noticed the handedness of near-horizon spirals could constrain spin and developed two techniques to measure the spiral handedness, but did not measure their pitch angle. Our work complements R. Sant et al. (in preparation), who also study near-horizon spiral structure using a different methodology and find consistent results.

The remainder of this paper is structured as follows.  In Section \ref{sec:method} we define the pitch angle $\psi$ and provide a validated method for measuring it in a fiducial GRMHD model. In Section \ref{sec:results}, we show that $\psi$ depends on spin. In Section \ref{sec:Discussion}, we discuss the physical interpretation, consider the feasibility of $\psi$ measurements in current and future data, and present model uncertainties. We conclude in Section \ref{sec:conclude}.

\section{Defining and Measuring Pitch Angle}\label{sec:method}
\subsection{GRMHD Models}
To interpret future EHT movies, it is necessary to start with synthetic observations based on models.  We use models similar to those used in previous analysis of EHT data \citep{M87PaperV, M87PaperVIII, SgrAPaperV}.  

The models used here are part of the \enquote{Illinois v5} \patoka GRMHD library \citep[][A. Joshi et al. in preparation, N. Bowden et al. in preparation]{Wong_2022}. The GRMHD simulations were run with the open-source, GPU-enabled general relativistic magnetohydrodynamics (GRMHD) code \kharma \citep{Prather_kharma_2024}. 
The \kharma fluid simulations are defined by two parameters. The first parameter is the dimensionless black hole spin $\spin\in [-1 < \spin < 1]$, where $\spin > 0$ denotes a prograde system (aligned black hole spin and disk angular momentum vectors) and $\spin < 0$ denotes retrograde systems (antiparallel black hole spin and disk angular momentum vectors). Tilted systems are not considered in this work. The second parameter is the magnetization mode, which takes two states: MAD \citep[strongly magnetized;][]{mad1_1974, mad2_2003, mad3_2003} or SANE \citep[weakly magnetized;][]{sane1_2012, sane2_2013}.\footnote{The physical, continuous parameter is the magnitude of the magnetic flux through one hemisphere of the black hole $\Phi$, or the dimensionless magnetic flux $\phi \equiv \Phi/(r_g^2 c\dot{M})^{1/2}$ for accretion rate $\dot{M}$; MAD models have $\phi \sim 16$ in Lorentz--Heaviside units, and $\sim 1$ for our SANE models.} The bulk of this work focuses on MAD models, since  M87* and Sgr A* are both favored to be MADs based on the linear polarization fraction and the magnitude of the 2nd mode $|\beta_2|$ of Fourier decompositions to the linear polarization structure \citep[][]{M87PaperVIII, SgrAPaperVIII}. 
SANEs are discussed in Appendix \ref{sec:appendix_othermodels}.

GRMHD fluid simulations are imaged using the radiative transfer code \ipole \citep{Mosc_2018}. 
The radiative transfer has three parameters. First is the viewing inclination $i$ of the flow, defined as the angle between the accretion flow angular momentum vector and the line of sight ($i=0$ is face-on above a counter-clockwise rotating disk, $i=90^\circ$ is edge-on, and $i=180^\circ$ is face-on below a clockwise rotating disk). Finally, there are electron temperature parameters $\rhigh$ and $\rlow$ \citep{Wong_2022}, which are inversely proportional to electron temperature.\footnote{The electron temperature $T_e$ is defined relative to the ion temperature $T_i$ with the following ratio: $T_i/T_e = \rhigh \,\tilde{\beta}^2/(1+\tilde{\beta}^2) + \rlow \,1/(1+\tilde{\beta}^2)$ for $\tilde{\beta}=\beta/\beta_{\text{crit}}$, plasma $\beta \equiv P_{gas}/P_{mag}$, gas pressure $P_{gas}$, magnetic pressure $P_{mag}$), and $\beta_{\text{crit}}=1$. }  

We focus on Sgr A* models, although M87* models are briefly discussed. In the v5 library, models are imaged from $t=20000$ to $50000 \, t_g$, with a cadence of $dt=5\,t_g$ and a pixel angular resolution of $dx=0.5\,\mu$as. Models from the earlier v3 library \citep{Dhruv_grmhd_survey_2024}, which uses a different adiabatic index \citep{Gammie_2025}, are discussed in the Appendix.

\subsection{The Pitch Angle} \label{subsec:pitch}

We define the pitch angle $\psi$ as the angle between the (outward-directed) tangent to a spiral arm and 
the outward-directed radial vector. 
Small $\psi>0$ corresponds to a spiral loosely winding toward $\Delta\PA>0$ (counterclockwise). Small $\psi<0$ corresponds to a spiral loosely winding toward $\Delta\PA<0$ (clockwise). 
The \enquote{spiral} becomes entirely radial at $\psi = 0^\circ$ and entirely azimuthal at  $\psi = \pm 90^\circ$. 

Notice that our definition for pitch angle  $\psi$ has symmetry about $\pm 180^\circ$ and is bounded by $\psi\in [-90^\circ, 90^\circ]$, with the useful property that the sign of $\psi$ conveys the handedness of the spiral. This differs from the usual \cite{binney_tremaine} definition, who define the pitch angle $\alpha$ as the magnitude of the complementary angle of $\psi$: $\alpha=90^\circ - |\psi|$ (so $\alpha$ graphically corresponds to the smallest angle between the spiral's tangent and a circle). 

The total intensity pitch angle invokes comparison with $\angle\beta_2$, the pitch angle of the electric vector polarization angle (EVPA) \citep{Palumbo_2020}. While $\angle\beta_2$ is not defined using the same convention as $\psi$, the typical $\angle\beta_2$ convention does favor radial EVPAs at $\angle\beta_2=0$, counterclockwise EVPAs at $\angle\beta_2>0$, and clockwise EVPAs at $\angle\beta_2<0$, like our convention. Discounting Faraday rotation, EVPAs fall perpendicular to magnetic field lines. These field lines are expected to align with spiral features, as can be seen in our GRMHD simulations and observed in spiral galaxies \citep{Beck_2000}. Thus, under the same sign conventions and ignoring Farday rotation, $\angle\beta_2$ and $\psi$ are expected to have opposite sign. The pitch angle might also be compared to the inclination. Assuming the spirals are trailing and do not change sign across radius, $\psi < 0$ corresponds to accretion flow angular momentum vector directed toward the observer ($\cos i > 0$), while $\psi > 0$ corresponds to accretion flow angular momentum vector directed away from the observer ($\cos i < 0$). Thus, $\psi$ and $\cos i$ should also have opposite sign.

\subsection{Cylinder Plots}\label{subsec:cyl}

The movie data takes the form of brightness temperature $T_b$ as a function of image plane position and time.  We make no attempt at image reconstruction using the EHT's restricted $\uv$ sampling or observational noise. Instead, we mimic EHT observations by convolving each frame with a Gaussian of full width at half maximum (FWHM) of $20\,\mu$as, matching the EHT's nominal resolution. To mimic BHEX observations, we instead use FWHM$=6\,\mu$as.

Following \cite{Conroy_2023} and \cite{Conroy_2026a}, we perform a coordinate transformation. We define a cylinder plot $T_b=T_b(\PA, t; b)$, which is a function of position angle $\PA$ on the sky (measured counterclockwise from North) and time $t$, and depends on the image-plane angular radius (or impact parameter) $b$ where the cylinder is measured.  In this work, we set the origin ($b = 0$) at the center of the critical curve, which is known analytically for synthetic images. As in \cite{Conroy_2023}, we measure cylinders across radii relative to the angular radius $b_0\equiv\sqrt{27}GM/(c^{2}D)$ for distance $D$ to the source \citep[which coincides with the critical curve to first order in spin;][]{Gralla_2020}. In observations, the ring location and radius will need to be measured using a feature extraction algorithms. Cylinder plot properties are expected to be insensitive to small errors in their location and radius \citep{Conroy_2023}. 

\begin{figure}[t]
    \centering
    \includegraphics[width=1.0\textwidth]{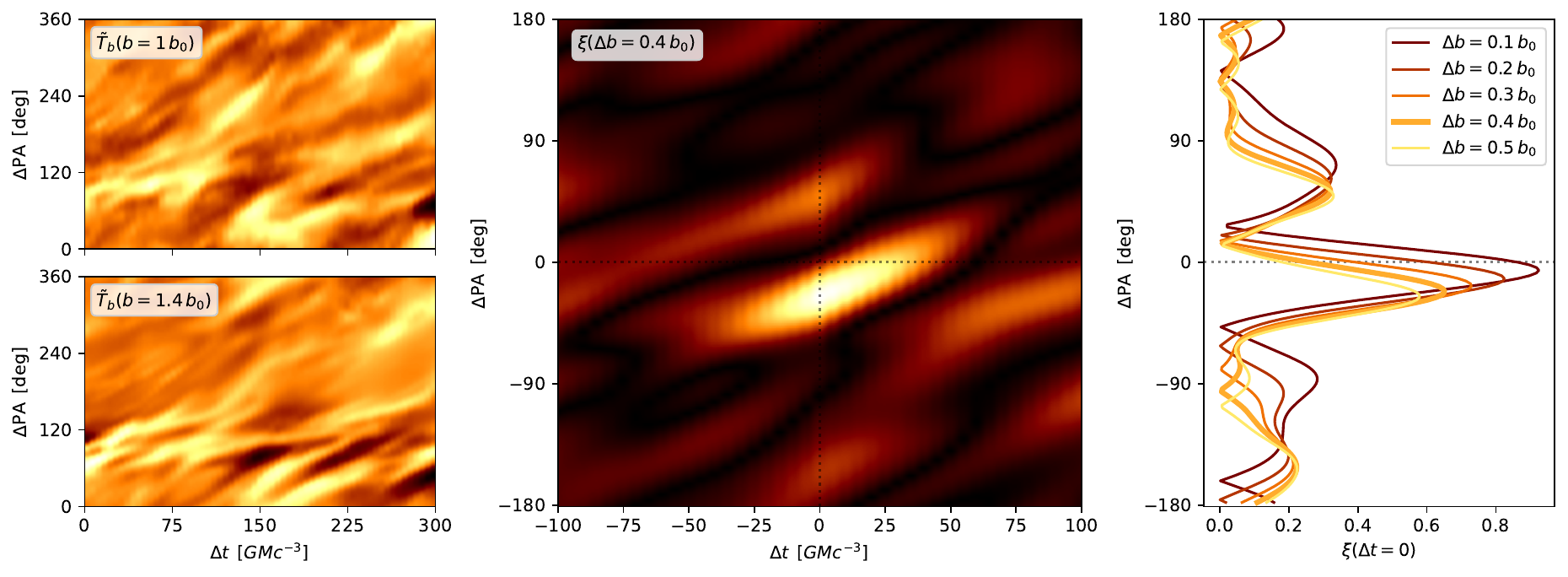}
    \caption{Left: Example cylinder plots showing normalized brightness temperature $\tilde{T}$ as a function of $(\PA,t)$ from the same fiducial model (MAD, $\spin=0.5$, $i=30^\circ$, $\rhigh=160$), measured at an angular radius of $b=1\,b_0$ (top left) and $1.4\,b_0$ (bottom left) with BHEX angular resolution.
    Middle: The cross-correlation of these two cylinders, with $b_{inner} = 1\,b_0$ and $\Delta b= 0.4\,b_0$. The central correlated peak falls along $\Delta\PA<0$, suggesting $\psi < 0$. Right: The measured cross correlation along the $\Delta t=0$ axis, as a function of $\Delta \PA$ and $\Delta r$. Notice as $\Delta r$ increases, the peak falls along decreasing $\Delta \PA$, suggesting $\psi < 0$ across this range of radii. 
    }
    \label{fig:autocor}
\end{figure}

The cylinder plot is then normalized by mean-subtracting along each $\PA$ slice then each $t$ slice, yielding $\tilde{T}_b(\PA, t; b)$. Unlike previous work \citep{Conroy_2023, Conroy_2026a}, we do not use the logarithm during normalization, as we will measure $\psi$ across a range of radii where emission and brightness asymmetry are low. The left panels of 
Figure \ref{fig:autocor} show normalized cylinder plots for a fiducial model at two radii.

\subsection{Measuring the Pitch Angle from Correlations}

EHT data has limited spatial resolution: in M87* and Sgr A*, the photon ring has diameter $\approx 50\,\mu$as compared to the nominal EHT beam $\approx 20\mu$as.  
It is thus difficult to visually identify, in a single low-resolution image, the spiral features that are easily visible in the higher resolution GRMHD snapshot (see Figure \ref{fig:frame}).  
If multiple images or a time sequence of snapshots is available, however, then correlations in the noisy, low-resolution data contain enough information to extract the sign of the pitch angle and constrain the magnitude.

Our main tool is the dimensionless cross-correlation
\begin{equation}
    \xi(b_{inner}, \Delta b, \Delta \PA) \equiv 
    \frac{\langle \, \tilde{T}_b(\PA, t; b_{inner}) \, \tilde{T}_b(\PA + \Delta\PA, t; b_{inner} + \Delta b) \, \rangle}{\langle \, \tilde{T}_b(b_{inner})^2 \, \rangle^{0.5} \, \langle \,\tilde{T}_b(b_{inner} + \Delta b)^2\,\rangle^{0.5}}
\end{equation}
where $\<\>$ indicates an average over $\PA$ and $t$. 

We correlate two cylinders measured along angular radius $b_{inner}$ and $b_{inner} +\Delta b$ to find $\xi=\xi(\Delta t, \Delta \PA; b_{inner}, \Delta b)$. An example cross-correlation is shown in the middle panel of Figure \ref{fig:autocor}. Notice that the cross-correlation exhibits a strongly-correlated peak ridge near the center, surrounded by low-amplitude noise. The slope of this cross-correlation corresponds to the pattern speed of features that cross both radii \citep{Conroy_2023, Conroy_2026a}. 
As we will show, the vertical shift of the peak corresponds to the pitch angle. 

We measure the vertical shift $\Delta \PA_{\text{max}}$ of the peak cross-correlation along the $\Delta t=0$ axis, which corresponds to correlated features that simultaneously span both cylinders.
We fit a second-order polynomial to the peak region, taking $\pm5$ points around the maximum $\xi(\Delta t = 0)$ and shrinking the width if there is a local minima within this range. The width was set based on validation tests in Section \ref{sec:test}. The location of the peak implies that the cylinder plot at $b_{inner}$ is most similar to the cylinder plot at $b_{inner}+\Delta b$ when rotated by $\Delta\PA_{\text{max}}$. Examples of $\xi(\Delta t=0)$ across different $\Delta b$ are shown in the right panel of Figure \ref{fig:autocor}. For constant $b_{inner}$ and increasing  $\Delta b$, the magnitude of the correlation peak decreases and $\Delta\PA_{\text{max}}$ decreases (as the spirals continue winding in the same direction). 

We estimate the pitch angle from $\Delta \PA_{\text{max}}$ using a logarithmic spiral model, which is a curve of constant phase (i.e. constant pitch angle). In on-sky polar coordinates with angular radius $b$ and position angle $\PA$, these are typically shown as $b=a \, e^{\cot(\psi)\, (\PA - \PA_0)}$ in exponential form or $\PA-\tan(\psi)\, \ln (b/a) = \PA_0$ in logarithmic form. Here, $a$ scales the spiral, $\psi$ is the pitch angle, and $\PA_0$ sets the initial angle of the spiral. We can measure $\psi$ in models using $\Delta\PA_{max}$ from the correlation:
\begin{equation}
    \tan \psi \equiv \frac{d\PA}{d \ln b} \approx \frac{\Delta\PA_{max}}{\Delta \ln b}= \frac{\Delta\PA_{max}}{\ln(b_{inner} + \Delta b) - \ln(b_{inner})}.
\end{equation}
Then, the pitch angle is 
\begin{equation}
    \psi \equiv \tan^{-1}(\frac{d\PA}{d \ln b} ). 
\end{equation}
For a logarithmic spiral, $\psi$ is constant, independent of $b$. In models, spirals may vary and need not be perfectly logarithmic, so we measure a mean $\psi$ as a function of $b$, as well as model parameters such as inclination and spin.

For our fiducial model, the resulting pitch angle is depicted  as the dashed lines in Figure \ref{fig:frame} (measured assuming BHEX resolution in a movie of duration $300 \,t_g$). Dividing the full duration of the fiducial model into videos of $300 \,t_g$, then measuring $\psi$ across $b_{inner}\in[0.5,1.9]\,b_0$ with $\Delta b=0.2 \,b_0$, we find a circular mean and standard deviation of $\psi=-46^\circ \pm 09^\circ$.
Repeating the measurement at EHT resolution, we find $\psi=-23^\circ \pm 11^\circ$. The mean and standard deviation at EHT resolution increase only slightly with increasing $\Delta b$. 
At EHT resolution, the magnitude of $\psi$ in this model is biased to be more radial, as faint azimuthal structure is blurred to form the bright ring. Nevertheless, even at EHT resolution, the sign of the observed $\psi$ is still correctly recovered: $\psi_{\text{obs}}<0$. 

\subsection{Validation with Logarithmic Spirals}\label{sec:test}
To verify the methodology, we generated movies containing random logarithmic spirals superimposed on an asymmetric ring background \citep[an $m$-ring with asymmetry $a_1=0.5$, mean amplitude $0.5$, and Gaussian thickness $\sigma_r=0.5\,b_0$, following the conventions of][]{Bernshteyn_2026, Conroy_2026_PA}. The spirals were randomly generated in initial position angle, number (mean $n=10$), and azimuthal thickness (mean $\Delta \PA=5^\circ$). Each spiral's amplitude evolved in time following a Gaussian curve, with a randomly generated correlation time (mean $100\,t_g$) and peak amplitude (mean $0.25$). The spirals had an analytically set pitch angle and rotated with an analytically set pattern speed ($|\Omega_p|=1$ deg per $t_g$). 
We performed $>1000$ tests with duration $300 \, t_g$ and cadence $dt=5\,t_g$, then compared the true pitch angle $\psi_{\text{true}}$ with the measured pitch angle $\psi_{\text{obs}}$.

Measurements become less accurate with increasing $|\psi_{\text{true}}|$, tending to slightly overestimate $|\psi|$. Past $\psi_{\text{true}}\geq \tan^{-1}(\pi/\Delta \ln b)$ (which is $\approx87^\circ$ for $\Delta b=0.2\,b_0$ and $\approx81^\circ$ for $\Delta b=0.6\,b_0$ at $b_{inner}=1b_0$), spirals become too tightly wound to recover using cylinder cross-correlation, as they undergo more than a half-turn between the two cylinders. For $|\psi_{\text{true}}| < 80^\circ$ and $\Delta b=0.2\,b_0$, 
the above method recovers a root mean squared error (RMSE) of $0.5^\circ$ at BHEX angular resolution, and the majority of samples have error $<1^\circ$,
suggesting $6 \,\mu$as angular resolution is sufficient to accurately recover $\psi$. 
Blurring to EHT angular resolution produces RMSE=$6.2^\circ$, although increasing $\Delta b$ to $0.6\,b_0$ reduces the RMSE to $4.4^\circ$. The EHT thus may be able to measure the sign and constrain the magnitude of $\psi$. 

The accuracy improves slightly for higher cylinder grid resolution (we use a default of $d\theta=2^\circ$), wider spirals (stronger pitch angle signal), and higher spiral sample size (e.g. longer video duration).
Rather than exhibiting a single $\psi_{\text{true}}$ that spans the frame, horizon-scale data may exhibit a range of $\psi$ with a characteristic mean as a function of $\PA$ and $b$ and a radial correlation length. The introduction of varying $\psi$ and limited correlation length could both increase error in pitch angle measurements.

The above methodology can also be applied to static images. Comparable measurement accuracy to the above video tests can be reached with only 2-3 images, provided the random spirals are independent with a constant $m$-ring background. This is expected, as the videos span a comparable number of correlation times without including imaging error. Analyzing $\psi$ in a single image requires modifying the methodology in Section \ref{subsec:cyl}: since the cylinders now contain only a single timeslice, we cannot normalize the cylinders in time. This increases error, as average non-spiral structure are included in the correlation. We then recover an RMSE of $10.6^\circ$ at BHEX resolution ($\Delta b=0.2\,b_0$) and $20.1^\circ$ at EHT resolution ($\Delta b=0.6\,b_0$).

\section{Pitch Angle Results}\label{sec:results}

\subsection{Spin Dependence}
\begin{figure}[ht]
    \centering
    \includegraphics[width=1.0\textwidth]{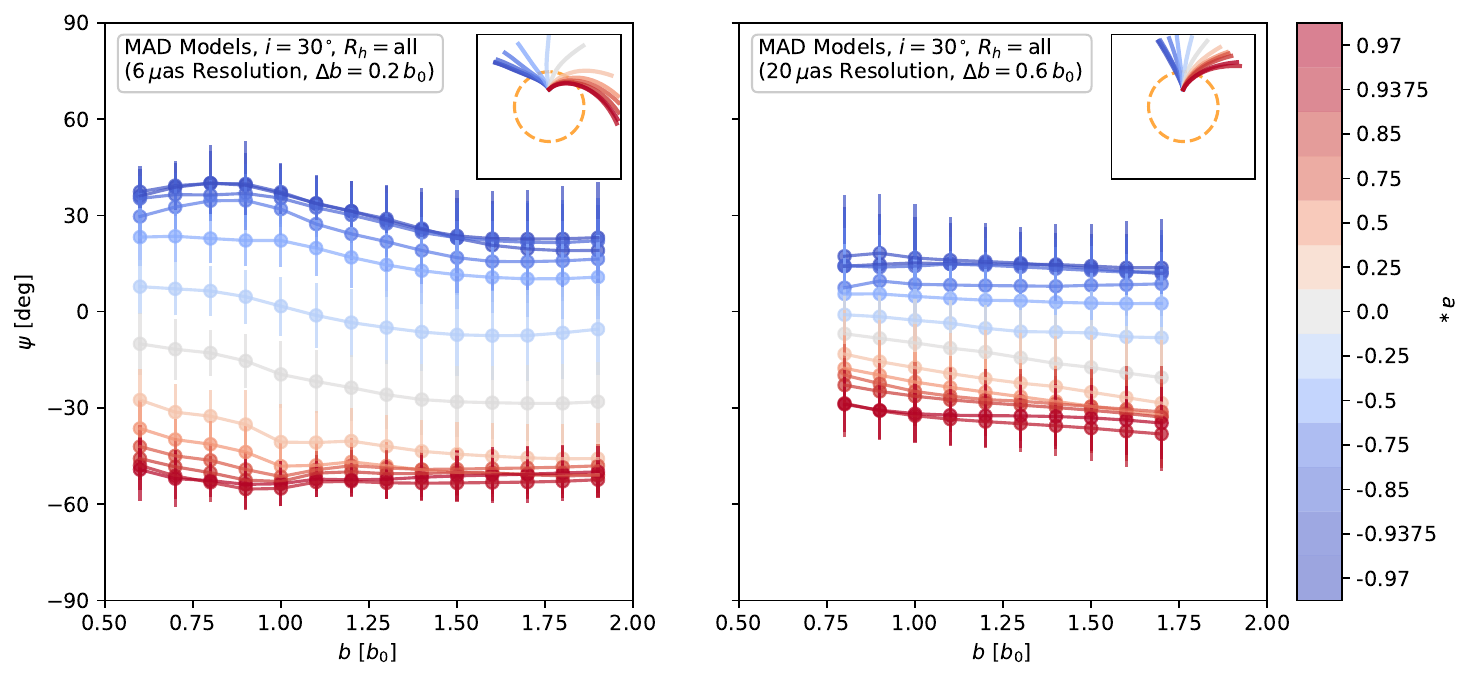}
    \caption{Here, we plot radial curves of $\psi$ measured in $300\,t_g$ movies of MAD models at $i=30^\circ$, across spin, marginalizing over $\rhigh$. The $y$-axis shows the measured pitch angle $\psi$ with error bars illustrating $1\sigma$ circular standard deviations, while the $x$-axis shows the midpoint angular radii $b$ around which the pitch angle was measured. Left: $\psi$ is measured at BHEX resolution with an angular separation of $\Delta b=0.2\,b_0$. Right: $\psi$ is measured at EHT resolution with an angular separation of $\Delta b=0.6\,b_0$. Insets in the top right of each subplot illustrate the mean measured $\psi$ across spin (designated by color) relative to the critical curve, which is shown as a dashed orange ring. Radial spirals with $\psi=0$ appear as vertical lines, while non-zero $\psi$ corresponds to spirals that wind around the ring.    }
    \label{fig:radial_spin}
\end{figure}

Applying the method of Section \ref{sec:method} to the full v5 MAD GRMHD library, we find pitch angles exhibit a clear spin dependence. Figure \ref{fig:radial_spin} illustrates the circular mean and circular standard deviation of $\psi$ across $\spin$ and angular radius $b$, for $i=30^\circ$, marginalizing over all other parameters (time, $\rhigh$, $\rlow$). In the left panel, we plot results at BHEX resolution. In the right panel, we plot results at EHT resolution. 

The sign and magnitude of $\psi$ exhibits a spin dependence. First, consider the sign. Near the horizon, $\psi$ exhibits the opposite sign as $\spin$. In our convention, the direction of the angular momentum vector on-sky is held constant across $\spin$ (pointed toward the observer at $i=30^\circ$), while the spin vector flips direction in  retrograde models (pointed away from the observer for $\spin<0$). Thus, near the horizon, $\psi$ is consistent with spirals that are trailing relative to the direction of the black hole spin.

The magnitude of $\psi$ increases as $|\spin|$ increases. That is, spirals become more tightly wound with increasing spin magnitude. Pitch angle is minimized (features appear most radial) near $\spin=0$ and $=-0.25$. 
Thus, if the black hole is sufficiently face-on and if $\spin$ can be shown to be non-zero \citep[as may be the case in M87*, based on the brightness asymmetry analysis;][]{Bernshteyn_2026, Conroy_2026_PA}, then a measurement of the sign of $\psi$ near the horizon enables a constraint on whether the black hole is prograde or retrograde. 

A detection of $\psi$ changing sign across radius also acts as a robust detection for a retrograde system. This is consistent with prior analysis \citep{Ricarte_2022}, who note spiral handedness flips near retrograde black holes due to frame dragging. 
The $\psi$ behavior we see is consistent with frame dragging; additional physical interpretation is discussed in Section \ref{sec:cause}. 

Notably, while the magnitude of $\psi$ decreases and the standard deviations increase for videos blurred to EHT resolution, these trends are still present. This suggests there may be a signal or spin in current EHT data, provided EHT has the required sensitivity and $\uv$ coverage (see also Section \ref{subsec:feasibility}). 

Pitch angle also depends on inclination, as discussed in detail in Appendix \ref{sec:appendix_othermodels}.  The direction of the spiral flips going from a given inclination to the supplementary inclination angle (from $i$ to $180-i$), as expected by symmetry. Since the spirals are trailing, the sign of $\psi$ sufficiently far from the horizon (for retrograde models, or across all radii for prograde models) indicates the direction of the disk angular momentum vector on the sky and thus constrains $i$. This would provide a consistency check on other observables that also constrain $i$, such as the pattern speed $\Omega_p$ \citep{Conroy_2023} and the linear polarization pitch angle $\angle\beta_2$ \citep{Palumbo_2020}.

In Figure \ref{fig:radial_spin}, we
marginalize over $\rhigh$. We find $\psi$ in MAD, face-on models is relatively insensitive to the electron temperature parameter. Mean values of $\psi$ vary by no more than several degrees across differing $\rhigh$, when measured at BHEX resolution with $\Delta b=0.2\,b_0$ and EHT resolution with $\Delta b=0.6\,b_0$. This is expected: in MAD models, the emission region is relatively insensitive to electron temperature values \citep{M87PaperV, Wong_2022}.

Consideration of $\psi$ in weakly magnetized SANE models, at varying inclination, and in earlier model libraries can be found in Appendix \ref{sec:appendix_othermodels}. In short: we find broad consistency in the spin-dependence of $\psi$ across other models. 

\section{Discussion}\label{sec:Discussion}

\subsection{Physical Origin}\label{sec:cause}

\cite{Ricarte_2022} interpreted the flip in the sign of $\psi$ as a signature of near-horizon frame dragging. Our results are consistent with expectations based on frame dragging. Near the horizon, the sign of $\psi$ depends on the direction of frame dragging, and the magnitude of $\psi$ changes with the magnitude of spin. 

\begin{figure}[ht]
    \centering
    \includegraphics[width=0.5\textwidth]{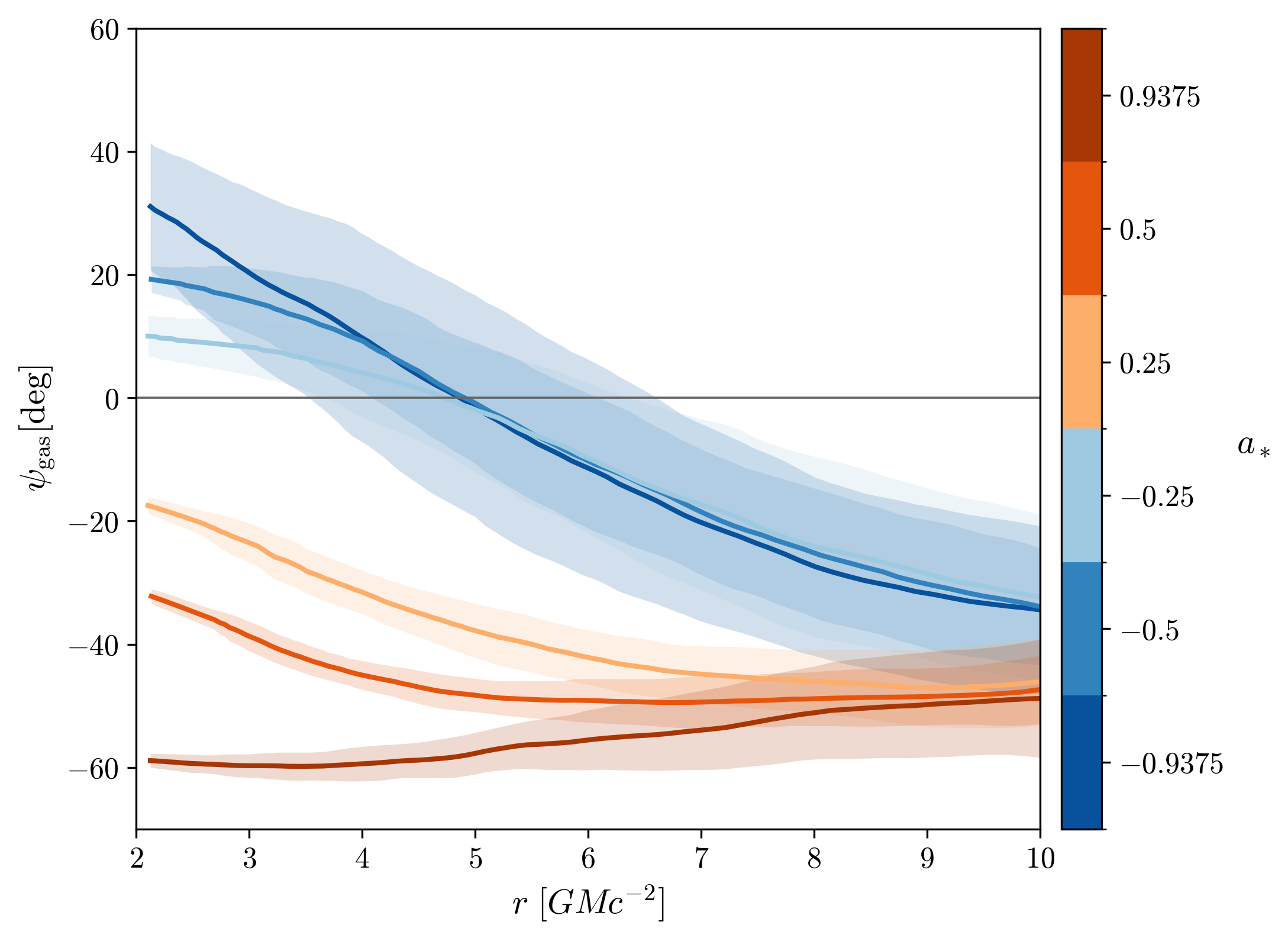}
    \caption{Circular mean values of $\psi_{\text{gas}}$ (with $1\sigma$ circular standard deviation contours) measured in $1000\,t_g$ movies of the gas pressure $p_g$ within $10^\circ$ of the midplane, for several MAD models. Error bars correspond to $1\sigma$ standard deviations.}
    \label{fig:pressure}
\end{figure}

In preceding work, we measured slow rotation in videos of GRMHD models using correlations, then interpreted the image-domain dynamics as being dominated by pressure fluctuations \citep{Conroy_2023, Conroy_2026a}. If true, we might expect $\psi$ to be dominated by pressure fluctuations here as well. In Figure \ref{fig:pressure}, we measure the pitch angle $\psi_{\text{gas}}$ of gas pressure $p_g$ fluctuations near the midplane (using the method described in Section \ref{sec:method}, with gas pressure cylinders as a function of azimuthal angle rather than brightness temperature cylinders as a function of image-domain $\PA$). The behavior of $\psi_{\text{gas}}$ is broadly consistent with the image-domain $\psi$. Near the horizon, the sign of $\psi_{\text{gas}}$ is set by frame dragging.
The magnitude changes across radius, as in the image domain.

Comparing retrograde values in Figure \ref{fig:pressure}
with Figure \ref{fig:radial_spin}, we find the midplane $\psi_{\text{gas}}$ changes sign at a smaller radius than $\psi$. This discrepancy is larger than what is expected from lensing alone. Off-midplane emission may be driving this effect. To test this hypothesis, we examined $\psi$ in a test model (MAD, $\spin=-0.5$, $i=30^\circ$, $\rhigh=160$), which was re-imaged while masking out emission beyond a $\Delta \theta$ range of the midplane \citep[see also the Appendix of][]{Conroy_2026a}. Masking out off-midplane emission, we found that $\psi$ changes sign at a lower radius, as anticipated. Appendix \ref{sec:appendix_othermodels} discusses $\psi$ in retrograde models from the earlier v3 GRMHD library, which has been hypothesized to exhibit less off-midplane emission than the updated v5  library \citep{Conroy_2026a}. Indeed, $\psi$ in retrograde v3 models change sign at a smaller radius than $\psi$ in retrograde v5 models. Thus, image domain measurements of $\psi$ are consistent with pressure fluctuations, after accounting for off-midplane emission. 

Beyond spiral pressure waves, other features may contribute to the measured pitch angle. GRMHD models predict the near-horizon accretion flow will exhibit both pressure fluctuations (including spiral sound waves and weak shocks) and entropy fluctuations (hot spots, e.g. flux tubes). While pressure fluctuations propagate through the disk with a phase velocity, entropy fluctuations advect with the fluid velocity. Entropy fluctuations exhibit some radial thickness $\Delta r$,  shear with the differentially rotating flow \citep{Tiede_2020, Porth_2021}, and so exhibit a pitch angle. 
Our measurement averages over all fluctuations, weighting them according to their contribution to nonaxisymmetric brightness fluctuations.
While pressure fluctuations may typically dominate, entropy fluctuations also contribute to the mean $\psi$ measured in GRMHD models, and the methodology proposed here could be used to measure $\psi$ in observations of these fluctuations.

\subsection{Feasibility in Current and Future Observations}\label{subsec:feasibility}
\begin{figure}[ht]
    \centering
    \includegraphics[width=0.48\textwidth]{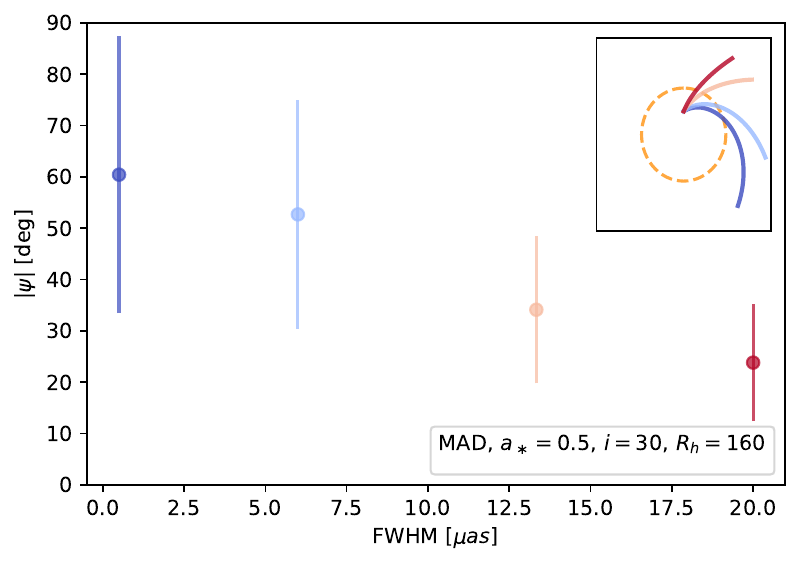}
    \hfill
    \includegraphics[width=0.48\textwidth]{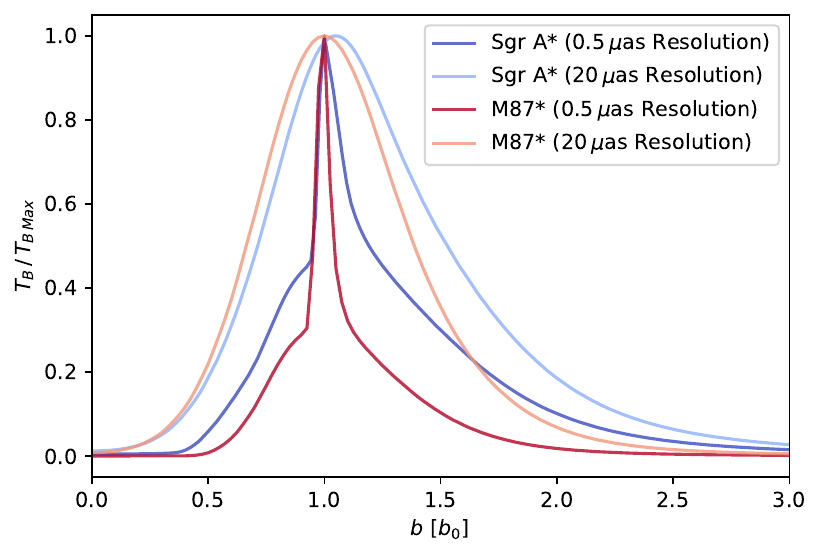}
    \caption{Left: $\psi$ in the fiducial model (MAD, $\spin=0.5$, $\rhigh=160$, $i=30^\circ$) versus the FWHM of the angular resolution blurring kernel. 
        Right: Radial profiles for two fiducial models ($i=30^\circ$ for Sgr A*, $i=17$ for M87*), illustrating the brightness temperature $T_B$ averaged across $\PA$ and time then normalized by the maximum value $T_{B\,max}$ as a function of $b$. }
    \label{fig:feasibility}
\end{figure}

A detection of $\psi$ in the image domain requires sufficient angular resolution. Section \ref{sec:results} suggests EHT has sufficient resolution to measure the sign of $\psi$ and even constrain the magnitude. Improvements to EHT's resolution, either through simultaneous higher-frequency observations or through space-ground interferometry with BHEX, should enable more precise measurements of the magnitude of $\psi$. 
This is emphasized in the left panel of Figure \ref{fig:feasibility}, where we plot the mean measured $\psi$ across  $b\in[0.5\,b_0, 2\,b_0]$ at $230$ GHz EHT resolution, future $345$ GHz EHT or ngEHT resolution (ignoring changes to optical depth at higher frequencies), BHEX resolution, and standard \ipole image resolution. 
Across all resolutions of this fiducial model, the mean $\psi$ is more than $1\sigma$ from zero.

We also require the dynamic range to measure structure across a range of radii. 
In the right panel of Figure \ref{fig:feasibility}, we plot the azimuthally and temporally averaged brightness temperature profile for the Sgr A* and M87* fiducial models as a function of radius. At EHT resolution, an angular radius of $b=1\pm0.25b_0$ has a brightness temperature of $\sim70\%$ the maximum of the temperature profile $T_{b\,max}$. By $2\,b_0$, the brightness temperature is $\sim0.2\, T_{b\,max}$ for Sgr A* and  $\sim0.07\, T_{b\,max}$ for M87*. Spiral features near the ring are brighter and may be easier for current or near-future EHT to extract. 
Nevertheless, the extraction of the pitch angle at EHT resolution may require the accurate reconstruction of subtle morphological features, faint structure, and extended structure.
This provides natural motivation for extensions to the EHT that substantially improve sensitivity, such as the ngEHT \citep{ Doeleman_2023_ngeht, Johnson_2023_ngeht}.

Finally,  $\psi$ measurements require sufficient $\uv$ coverage to enable either accurate image reconstruction (see \citealp{Ricarte_2022} for discussion of spirals in ngEHT image reconstructions) or $\psi$ extraction using the visibility data (see Appendix \ref{sec:appendix_log} for initial discussion). 

Beyond questions of feasibility, $\psi$ measurements depend on one hyperparameter, the $\Delta b$ separation between the cylinders. Current validation tests in Section \ref{sec:test} assume logarithmic spirals, while in truth we expect radially extended structure to exhibit some radial correlation length. Optimal extraction of $\psi$ in observational images might require calibrating $\Delta b$ as a function of the blurring kernel, radial correlation length, and other morphological parameters.
We leave in-depth study into $\psi$ recovery from VLBI observations to future work.

\subsection{Initial Analysis of $\psi$ in M87* and Sgr A*}
It is interesting to measure $\psi$ in current EHT image reconstructions, despite gaps in $\uv$ coverage introducing error in image domain $\psi$ measurements.

2017 M87* image reconstructions exhibit variation in $\psi$ across imaging pipelines and radius. The imaging methods {\tt DIFMAP} and {\tt SMILI} marginally favor clockwise spirals. In contrast, {\tt eht-imaging} favors counter-clockwise spirals for $b\gtrsim1\,b_0$. For all methods, $\psi$ measurements depend on extended structure which may be biased in 2017 reconstructions. More recent reconstructions of 2017 M87* using the imaging algorithm {\tt PRIMO} exhibit visually clear counterclockwise spirals. 
This would correspond to a clockwise pattern speed \citep{Conroy_2023} and thus a prograde system \citep{Conroy_2026_PA}. However, the PRIMO algorithm uses a training set of GRMHD images that only included prograde models. This motivates follow-up analysis of $\psi$ with {\tt PRIMO} and other imaging algorithms in subsequent years' data.

The 2017 method-averaged image of Sgr A* exhibits a slight preference for counterclockwise spirals near the bright ring. This would be consistent with the mild preference for $i>90^\circ$ and prograde spins in 2017 analysis \citep{SgrAPaperV, SgrAPaperVIII}. However, this pitch angle constraint is tentative, as the sign of $\psi$ flips for off-ring structure (which may be poorly reconstructed or favor retrograde spins), because the 2017 reconstruction represents a time-averaged image across the April 7 observation rather than a snapshot image, and because analysis is best done within individual imaging pipelines with comprehensive validation tests for $\psi$ measurements. We leave rigorous analysis of pitch angles in Sgr A* to future work. 

\subsection{Model Uncertainty}
Our pitch angle results are model-dependent. Best-bet GRMHD models from our library successfully fit the majority of constraints leveraged by the EHT for both Sgr A* \citep{SgrAPaperV,SgrAPaperVIII} and M87* \citep{M87PaperV,M87PaperVIII,M87PaperIX,M87_2018_paperII}. This includes all EHT morphology-based constraints. However, the light-curve variability in Sgr A* models is known to be higher than in observations. This is expected to be driven by limitations and uncertainties in current GRMHD modeling. Discussions of these uncertainties can be found in \cite{Wong_2022, Bernshteyn_2026, Conroy_2026_PA}, and others. 

Key uncertainties in the GRMHD simulations include (1) the optimal treatment of boundary and initial conditions, (2) collisionless dynamics, and (3) thermodynamics and electron energization. Many of these uncertainties do not seem to change broad trends in $\psi$'s behavior. In Appendix \ref{sec:appendix_othermodels}, we analyze varying initial conditions that produce strongly magnetized MAD models and weakly magnetized SANE models, and varying thermodynamics treatments with ion adiabatic index $\gamma_i=4/3$ and $5/3$, and find similar trends as a function of spin, although the values of $\psi$ differ between models. Visual inspection of models with moderately tilted initial conditions, where the angular momentum of the disk is misaligned from the black hole spin vector, do not change the expected sign of $\psi$ across spin \citep[see][Figures 10, 11, C1, and C2]{Chatterjee_2020}. 

How do uncertainties in collisionless dynamics and electron energization influence pitch angle? 
The plasma accreting close to the event horizon in low-luminosity active galactic nuclei is collisionless, so kinetic effects may alter the amplitude and phase of the spiral waves in these systems (see, e.g., \citealp{Majeski_2026}, who find that viscous dissipation can strongly dampen spiral density waves low-collisionality shear flows), and hence the observed pitch angle. To test this, we measured $\psi_{\rm gas}$ in a weakly collisional extended-GRMHD model \citep{Dhruv_nonideal_2025} and in an otherwise identical ideal GRMHD control run. The two agree: over $r = 2$--$6\,GMc^{-2}$ the root-mean-square difference between the $\psi_{\rm gas}(r)$ profiles is $1.02^{\circ}$, comparable to the $1.2^{\circ}$ expected if the two profiles were identical, obtained by adding standard errors of the two measurements in quadrature. Both quantities decrease as the inverse square root of the number of time windows, suggesting the measured difference is consistent with statistical error alone. Visual inspection of models with more accurate two-temperature electron energization treatments also exhibit similar behavior in the magnitude and sign of $\psi$ to our results here (see Figure 6 of \citealp{salas_2024} and Figure 9 of \citealp{Chael_2025}).

Nonetheless, pitch angle may vary across other boundary conditions, initial conditions, fluid assumptions, and temperature treatments, motivating study of $\psi$ in other models. 

During radiative transfer, key uncertainties include (4) mass density scaling, (5) the so-called \enquote{fast-light} approximation, and (6) inclination uncertainty (in the case of M87*, where typical models assume $i=17^\circ$ or $163^\circ$). 

Small changes in the mass density scaling due to uncertainty in the source compact flux may have small effect on $\psi$; but large changes may be required for future sources, increasing optical thickness and jet emission and thus changing $\psi$ \citep[see][Figure 13]{Wong_2022}. This motivates $\psi$ analysis in models of future sources \citep{Zhang_2025, Akiyama_2026}. To assess the fast-light approximation (which performs radiative transfer for each model frame along a single timeslice, ignoring source changes during the light-crossing time), we consider two fiducial MAD models with $\spin=0.94$, $\rhigh=40$ at $i=163^\circ$ for M87* and $i=60^\circ$ for Sgr A*. At BHEX resolution sampling across $b\in[0.5,2\,b_0]$, we find the root mean squared difference is only $\sim2^\circ$ between the slow light and fast light models in both M87* and Sgr A*, suggesting the fast light assumption is not a significant source of error. Uncertainty in M87*'s inclination is also not expected to have a large influence on $\psi$, as discussed in Appendix \ref{sec:appendix_othermodels}.

Finally, $\psi$ measurements and subsequent spin-inference are subject to (7) measurement error and (8) statistical error. Reconstructions with ngEHT coverage may be sufficient to measure the sign of $\psi$ \citep{Ricarte_2022}. However, the accuracy with which $\psi$ can be recovered with varying $\uv$ coverage, and how $\uv$ coverage might bias the measurement, has yet to be investigated.

\section{Conclusion}\label{sec:conclude}
In this work, we conduct the first systematic survey of spiral pitch angles in GRMHD models of EHT sources. We develop a technique to measure the pitch angle, $\psi$, in both near-horizon images and movies of black holes. We validate the technique on logarithmic spirals, then analyze $\psi$ across model parameters. 

The magnitude of $\psi$ exhibits a clear spin dependence. For constant viewing inclination, the sign of $\psi$ near the horizon depends on the sign of spin.
Current EHT resolution is sufficient to constrain $\psi$ in M87* and Sgr A*, although an in-depth study of limitations due to sensitivity and $\uv$ coverage is left to future work. Far from the horizon (for retrograde models, and across all angular radii for prograde models), the sign of $\psi$ reveals the direction of the disk angular momentum vector, since the waves are trailing. This enables constraints on $i$ with static images, complementing other observables including the pattern speed \citep{Conroy_2023}, linear polarization pitch angle \citep{Chael_2026_beta2}, rotation signatures in the Stokes $Q-U$ plane \citep{Ricarte_2025}, and circular polarization morphology \citep{Joshi_2024}.

The dependence of $\psi$ on spin provides clear motivation for extending the current EHT. Proposed ground-based extensions, such as the ngEHT, could improve 
sensitivity, $\uv$ coverage, and angular resolution via $345$ GHz observations, all beneficial to the accurate extraction of $\psi$. Proposed space-based extensions, such as BHEX, could 
provide sufficient angular resolution to accurately measure $\psi$ in current EHT sources and constrain $\psi$ in other, less well-resolved sources.

Spin inference with $\psi$ would be improved using complementary observables, such as the brightness asymmetry \citep{Bernshteyn_2026, Conroy_2026_PA} and linear polarization pitch angle \citep[e.g.]{Palumbo_2020}. Combining the sign of $\psi$ with the sign of the pattern speed \citep{Conroy_2023, Conroy_2026a} in video reconstructions could provide joint constraints on the direction of both the spin vector and disk angular momentum vector, thus improving recovery of the sign of the spin. 
Spin is relevant to the launching of relativistic jets, the motion of material surrounding the supermassive black hole, and the growth history of the black hole itself. 
Thus, accurate spin inference should be a central goal for the EHT and future expansions in the years to come.

\section{Acknowledgments} \label{sec:Acknowledgment}

This work was supported by NSF grant AST 20-34306.  
N.C. is supported by the NASA Future Investigators in NASA Earth and Space Science and Technology (FINESST) program. This material is based upon work supported by the National Aeronautics and Space Administration under Grant No. 80NSSC24K1475 issued through the Science Mission Directorate.
This research used resources of the Oak Ridge Leadership Computing Facility at the Oak Ridge National Laboratory, which is supported by the Office of Science of the U.S. Department of Energy under Contract No. DE-AC05-00OR22725.  This research used resources of the Argonne Leadership Computing Facility, which is a DOE Office of Science User Facility supported under Contract DE-AC02-06CH11357.  This research was done using services provided by the OSG Consortium, which is supported by the National Science Foundation awards \#2030508 and \#1836650.  This research is part of the Delta research computing project, which is supported by the National Science Foundation (award OCI 2005572), and the State of Illinois. Delta is a joint effort of the University of Illinois at Urbana-Champaign and its National Center for Supercomputing Applications. This work was initiated in part at the Aspen Center for Physics, which is supported by National Science Foundation grant PHY-2210452.

This work was typeset using \LaTeX\ style openjournal v. 09/06/15. AI tools (GPT-5 and GPT-6) were used to assist in the development of the Validation scripts in Section \ref{sec:test}, in the formatting of various figures, and in the revision of the manuscript text. Responsibility for the final manuscript text lies entirely with the authors.

\vspace{5mm}

\appendix 

\section{Inclination Dependence, SANE Models, and the v3 Model Library}\label{sec:appendix_othermodels}
Here, we survey $\psi$ behavior across inclination, in weakly-magnetized SANE models, and in other earlier model libraries.

\begin{figure}[ht]
    \centering
    \includegraphics[width=1.0\textwidth]{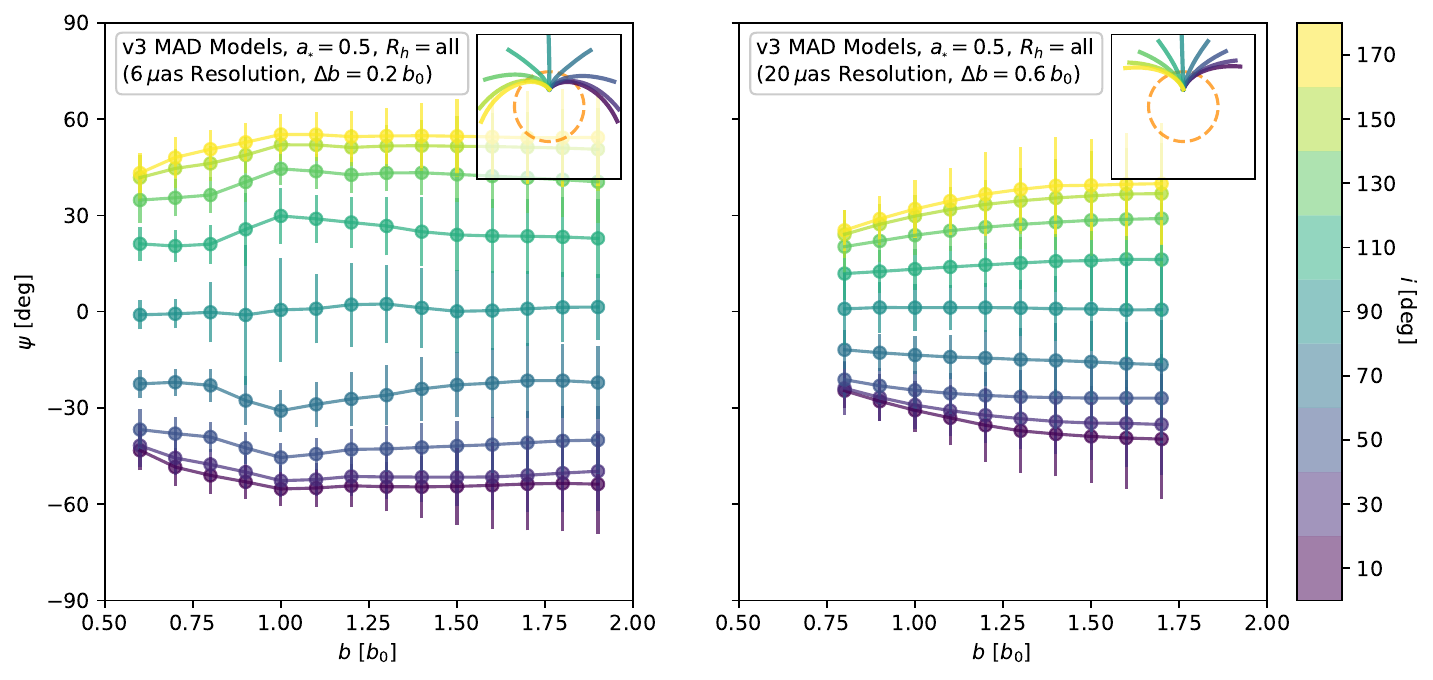}
    \caption{Here, we plot radial curves of $\psi$ as in Figure \ref{fig:radial_spin}, but now for MAD models at $\spin=0.5$ across inclination in the v3 library.}
    \label{fig:radial_incl}
\end{figure}

The inclination dependence shown in Figure \ref{fig:radial_incl} seems to approximately follow $\psi\propto\cos i$. As expected, $|\psi|$ is maximized when face-on. As expected, the apparent direction of the spiral flips when viewed face-on from above and face-on from below.
When edge-on, symmetry above and below the midplane drives a mean value of $\psi=0$. Changes in $\psi$ from $i=10^\circ$ to $30^\circ$ are small compared to the magnitude of $\psi$: notice these changes are also smaller than the $1\sigma$ standard deviations for each $i$ across radii. 
While these values of $\psi$ are calculated on Sgr A* models, we expect a similar inclination dependence to apply for $\psi$ in M87* models. There, model inclination is typically face-on at $i=17^\circ$ or $163^\circ$ \citep[based on assumed alignment with the large-scale jet axis, with standard deviation uncertainty of $\sim3^\circ$ to $4^\circ$;][]{Mertens_jet_2016, walker_jet_2018}. Thus, uncertainty in M87*'s inclination is unlikely to dominate measurements of $\psi$ or $\psi$-based spin inference.

\begin{figure}[ht]
    \centering
    \includegraphics[width=1.0\textwidth]{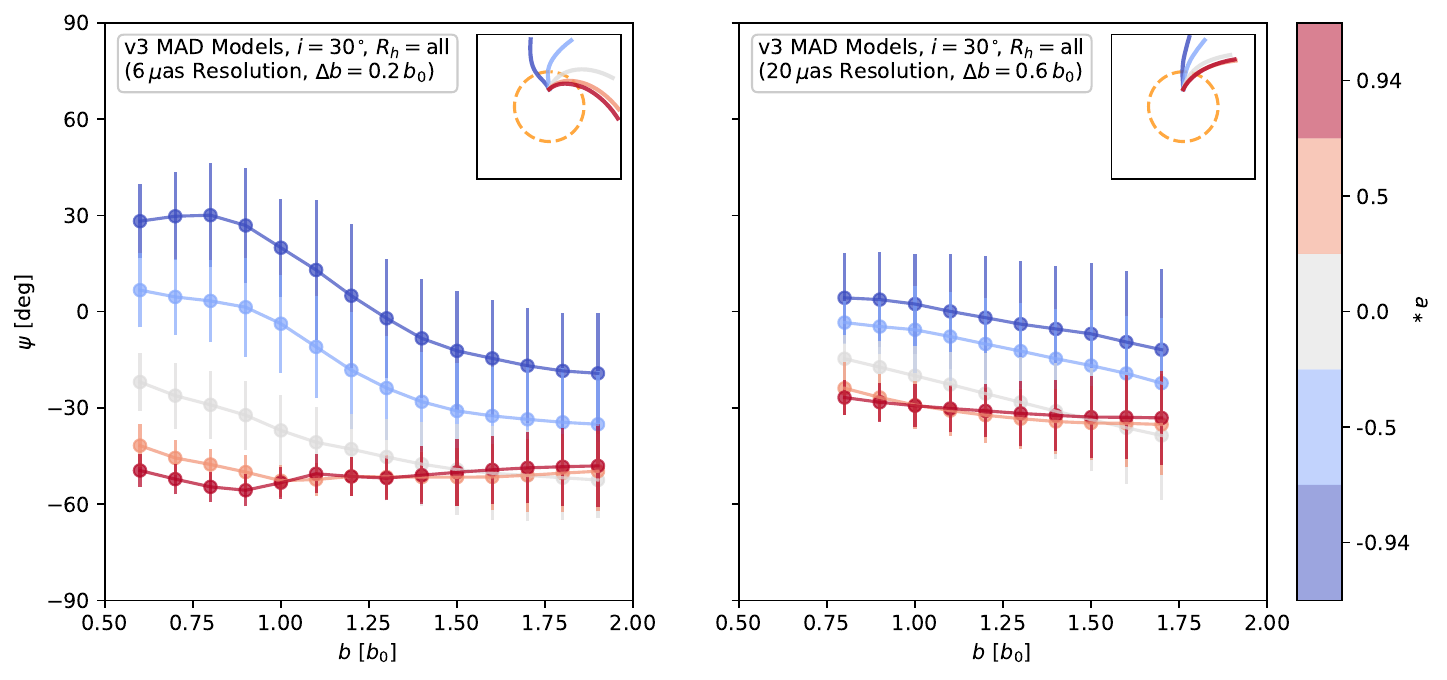}
    \caption{Here, we plot radial curves of $\psi$ as in Figure \ref{fig:radial_spin}, but now for MAD models at $i=30^\circ$ across spin in the v3 library.}
    \label{fig:radial_spin_v3}
\end{figure}

\begin{figure}[ht]
    \centering
    \includegraphics[width=1.0\textwidth]{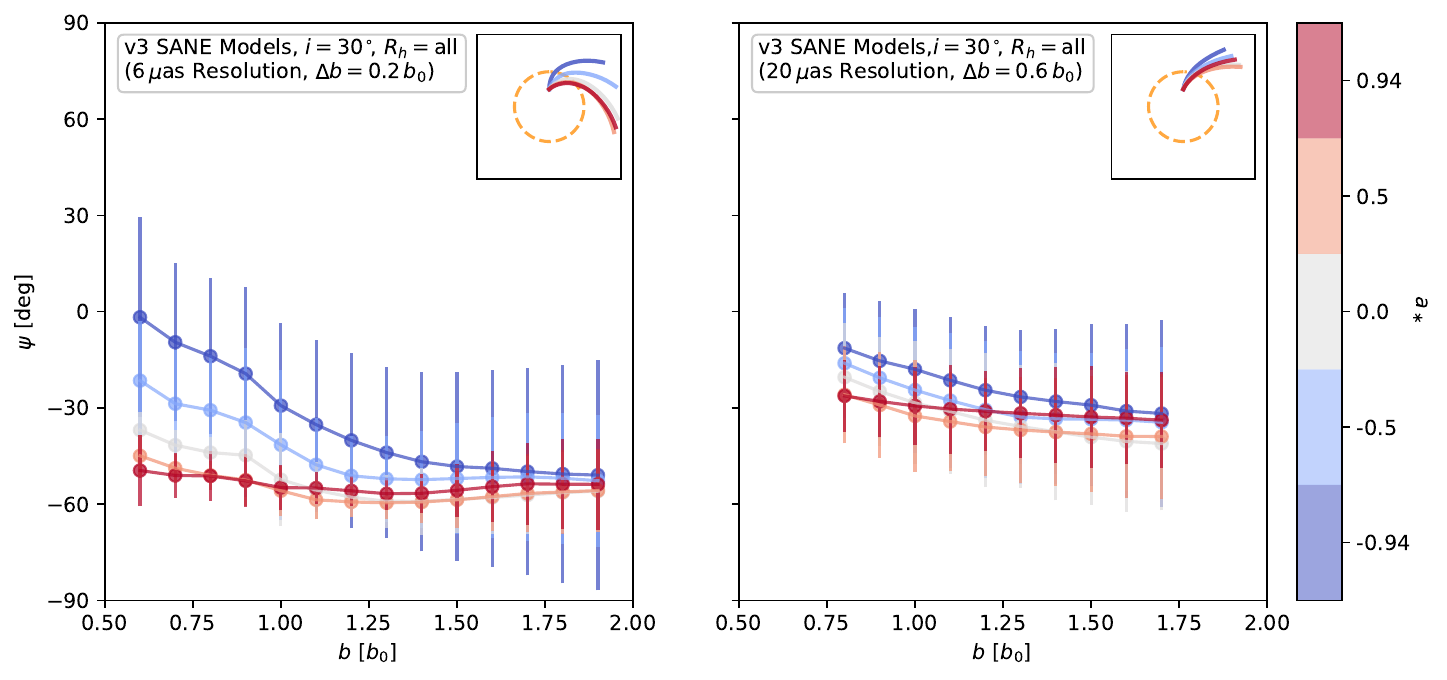}
    \caption{Here, we plot radial curves of $\psi$ as in Figure \ref{fig:radial_spin}, but now for SANE models at $i=30^\circ$ across spin in the v3 library.}
    \label{fig:radial_spin_SANE}
\end{figure}

We find broadly consistent trends comparing $\psi$ in the strongly magnetized MAD models from the updated v5 library with $\psi$ in the earlier v3 GRMHD library (\citealp{Dhruv_grmhd_survey_2024}; run with a different adiabatic index of $\gamma_i=4/3$ rather than $5/3$, per \citealp{Gammie_2025}) or in weakly magnetized SANE models. Figure \ref{fig:radial_spin_v3} illustrates $\psi$ in v3 MADs and Figure \ref{fig:radial_spin_SANE} illustrates $\psi$ in v3 SANEs. Close to the horizon, the sign and magnitude of $\psi$ depends on spin. Further from the horizon (for retrograde models, and across all radii for prograde models), the sign of $\psi$ depends on the direction of the disk angular momentum vector. The magnitude of $\psi$ appears broadly consistent in high-spin prograde models. 

Still, the precise magnitude of $\psi$ differs between model sets, particularly as we spin down the black hole through $\spin=0$ to strongly retrograde values. This change is most apparent for SANEs, which require more strongly retrograde spins and smaller angular radii than v3 or v5 MADs before the sign of $\psi$ flips. This may be driven by the varying fluid velocity in these models. SANE models exhibit Keplerian rotation, requiring stronger retrograde frame dragging closer to the horizon before the sign of $\psi$ flips. MAD models from v3 exhibit slightly sub-Keplerian rotation, while MAD models from v5 slightly more sub-Keplerian rotation; thus, v5 MADs require less retrograde frame dragging further out in radius before $\psi$ flips sign. Other differences in SANE models include a larger standard deviation, in part driven by a stronger dependence on $\rhigh$ in SANEs. This is expected, as the emission region depends more strongly on $\rhigh$ in SANEs than in MADs \citep{M87PaperV}.

Small differences aside, broad consistency in the dependence of $\psi$ on $\spin$ and $i$ across models gives confidence that it may be a useful diagnostic for $\spin$ and $i$ in future observations.

\section{Logarithmic Spiral in the UV Domain}
\label{sec:appendix_log}
This work analyzes spiral pitch angles in GRMHD models in the image domain, but image domain measurements require accurate image reconstruction. Here, we motivate future extraction of $\psi$ from visibility data by starting with a simplified test problem: we show that image-domain logarithmic spirals retain the magnitude of their pitch angle but have a flipped helicity in the $\uv$ domain. 

As a model problem for the transformation of a spiral feature from the image domain with coordinates $(r,\theta)$ to the Fourier domain with coordinates $(k, \theta_k)$, consider a brightness profile with the following form:
\begin{equation}
    I(r,\theta) = I_0 \left(\frac{r}{r_0}\right)^{-n} \left(
    1 + \epsilon \cos(q \log (r/r_0) + m \theta + \phi)\right)
\end{equation}
where $q = -m \tan\psi$, $m$ is an integer, and $\psi$ is our pitch angle. For a given brightness profile, $I_0$, $r_0$, $\epsilon$, and $\phi$ are constants. 

The Fourier transform is here defined as
\begin{equation}
    \hat{I}(k,\theta_k) = \int I(r,\theta) \exp(i \mathbf{k} \cdot \mathbf{r}) \, r \, dr \, d\theta
\end{equation}
where the usual Fourier coordinates in radio astronomy are $u = k \cos\theta_k/(2\pi)$ and $v = k \sin\theta_k/(2\pi)$.   Notice that $\mathbf{k} \cdot \mathbf{r} = k r \cos (\theta - \theta_k)$.  

Expanding, we find 
\begin{equation}
    \hat{I}(k,\theta_k) = \int I_0 \left(\frac{r}{r_0}\right)^{-n} \left(
    1 + \epsilon \cos(q \log (r/r_0) + m \theta + \phi)\right) \exp(i k r \cos(\theta -\theta_k)) \, r \, dr \, d\theta.
\end{equation}
We can change variables to $R = r/r_0$ and $K = k r_0$, so that 
\begin{equation}
    \hat{I}(K,\theta_k) = (I_0 r_0^2) \int R^{-n} \left(
    1 + \epsilon \cos(q \log R + m \theta + \phi)\right) \exp(i K R \cos(\theta -\theta_k)) \, R \, dR \, d\theta.
\end{equation}
This can be separated into an axisymmetric piece (i.e. when $\epsilon \rightarrow 0$) and a non-axisymmetric piece. The axisymmetric piece becomes
\begin{equation}
    \hat{I}(K,\theta_k) = (I_0 r_0^2) \int R^{-n} \exp(i K R \cos(\theta -\theta_k)) \, R \, dR \, d\theta.
\end{equation}
The $\theta$ integral gives
\begin{equation}
    \hat{I}_0(K,\theta_k) = 2 \pi (I_0 r_0^2) \int R^{-n + 1} J_0(K R) \, dR, 
\end{equation}
and the radial integral gives 
\begin{equation}
    \hat{I}_0(K,\theta_k) = \hat{I}_0(K) = 
    \pi (I_0 r_0^2) (K/2)^{n - 2} \Gamma(1-n/2)/\Gamma(n/2).
\end{equation}
This becomes $2\pi I_0 r_0^2/K$ when $(n \rightarrow 1)$.

The nonaxisymmetric piece is 
\begin{equation}
    \hat{I}_m(K,\theta_k) = \epsilon (I_0 r_0^2) \int R^{-n} \cos(q \log R + m \theta + \phi) \exp(i K R \cos(\theta -\theta_k)) \, R \, dR \, d\theta.
\end{equation}
We can solve for the $\cos$ term to find 
\begin{equation}
    \cos(q \log R + m \theta + \phi) = \frac{1}{2}\left( e^{i q \log R + i m \theta + i \phi} + c.c. \right) = \frac{1}{2}\left( R^{i q} e^{i m \theta + i \phi} + c.c. \right)
\end{equation}
where the complex conjugate is denoted $c.c.$ here for brevity. Let's drop the complex conjugate for now (we can add it back in at the end by adding a term with the signs of $q, m, \phi$ inverted).  Substituting the $\cos$ term into $\hat{I}_m$, we find
\begin{equation}
    \hat{I}_m(K,\theta_k) = \frac{1}{2} \epsilon (I_0 r_0^2) \int R^{-n + i q} \exp(i m \theta + i \phi + i K R \cos(\theta -\theta_k)) \, R \, dR \, d\theta.
\end{equation}
The $\theta$ integral can be solved as 
\begin{equation}
    \int d \theta e^{i m \theta + i K R \cos (\theta - \theta_k)} = 2 \pi i^m e^{i m \Theta_k} J_m(K R)
\end{equation}
so the full integral becomes
\begin{equation}
    \hat{I}_m(K,\theta_k) = \pi \epsilon (I_0 r_0^2) i^m e^{i m \theta_k + i \phi} \int R^{-n + i q + 1} J_m(K R)\, dR.
\end{equation}
This is the Hankel transform of a power law:
\begin{equation}
    \int R^{s} J_m(K R)\, dR = 2^s K^{-1 - s} \Gamma((1 + m + s)/2)/\Gamma((1 + m - s)/2)
\end{equation}
provided $Re(s) < 1/2$ and $Re(m + s) > -1$, or $n > 1/2$ and $n < 2 + |m|$.  Putting it together, 
\begin{equation}
    \hat{I}_m(K,\theta_k) = \pi \epsilon (I_0 r_0^2) i^m e^{i m \theta_k + i \phi} 2^{1 - n + i q} K^{-2 + n - i q} \Gamma((2 + m - n + i q)/2)/\Gamma((m + n - i q)/2).
\end{equation}
This might be simplified with the aid of the complex conjugate.

Specializing to $m = 1$ and $n = 1$, 
\begin{equation}
    \hat{I}_1(K,\theta_k) = \pi \epsilon (I_0 r_0^2) i e^{i\theta_k + i\phi} 2^{i q} K^{-1 - i q} \Gamma(1 + i q/2)/\Gamma(1 - i q/2).
\end{equation}
Notice that the term $K^{-i q} = e^{-i q \log K}$ yields a logarithmic spiral, with pitch angle $\psi$ retained in the term $q = -m \tan\psi$. Notice the pitch angle magnitude is preserved in the $\uv$ domain, but the sign relative to $\theta_k$ is reversed, so that the helicity of the spiral has flipped.

This suggests the image domain pitch angle might be constrained by measuring the pitch angle in the visibility data.

\clearpage
\bibliographystyle{aasjournal}
\bibliography{main}

\end{document}